\documentclass[twocolumn,amssymb, nobibnotes, aps, prl,amsmath,mathtools, nofootinbib,pgfmath,pgffor,superscriptaddress]{revtex4-1}
\usepackage{graphicx}

\makeatletter
\def\qrr@split@result#1 #2\@qrr@split@result{\edef\erfInput{#1}\edef\erfResult{#2}}
\newcommand*{\gnuplotErf}[2][\jobname.eval]{%
    \immediate\write18{gnuplot -e "set print '#1'; print #2, erf(#2);"}%
    \everyeof{\noexpand}
    \edef\qrr@temp{\input{#1} }%
    \expandafter\qrr@split@result\qrr@temp\@qrr@split@result
}
\makeatother

\usepackage{caption,color}
\usepackage{subcaption}
\usepackage{multirow}

\usepackage{hyperref}
\hypersetup{colorlinks,citecolor=blue}

\begin{document}

\title[Echoes from the Abyss]{Echoes from the Abyss: The Holiday Edition!}

\author{Jahed Abedi}
\email{jahed$_$abedi@physics.sharif.ir}
\affiliation{Department of Physics, Sharif University of Technology, P.O. Box 11155-9161, Tehran, Iran}
\affiliation{School of Particles and Accelerators, Institute for Research in Fundamental Sciences (IPM), P.O. Box 19395-5531, Tehran, Iran}
\affiliation{Perimeter Institute for Theoretical Physics, 31 Caroline St. N., Waterloo, ON, N2L 2Y5, Canada}

\author{Hannah Dykaar}
\affiliation{Department of Physics, McGill University, 3600 rue University, Montreal, QC, H3A 2T8, Canada}
\affiliation{Department of Physics and Astronomy, University of Waterloo, Waterloo, ON, N2L 3G1, Canada}

\author{Niayesh Afshordi}
\email{nafshordi@pitp.ca}
\affiliation{Perimeter Institute for Theoretical Physics, 31 Caroline St. N., Waterloo, ON, N2L 2Y5, Canada}
\affiliation{Department of Physics and Astronomy, University of Waterloo, Waterloo, ON, N2L 3G1, Canada}

\begin{abstract}
In a recent paper \cite{Abedi:2016hgu}, we reported the results of the first search for echoes from Planck-scale modifications of general relativity near black hole event  horizons using the public data release by the Advanced LIGO gravitational wave observatory. While we found tentative evidence (at $\simeq	 3 \sigma$ level) for the presence of these echoes, our statistical methodology was challenged by Ashton, et al.  \cite{Ashton:2016xff}, just in time for the holidays! In this short note, we briefly address these criticisms, arguing that they either do not affect our conclusion or change its significance by $\lesssim 0.3\sigma$. The real test will be whether our finding can be reproduced by independent groups using independent methodologies (and ultimately more data). 
\end{abstract}

\maketitle


Recently we reported tentative evidence of Planck-scale structure at black hole event horizons \cite{Abedi:2016hgu} using the public data release for the three Advanced LIGO
black hole merger events GW150914, LVT151012 and GW151226. Accounting for the ``look elsewhere'' effect due to uncertainty in the echo template, we find tentative evidence for Planck-scale structure near black hole horizons at $2.9\sigma$ significance level (corresponding to false detection probability
of 1 in 270).

The key property of the signal that we searched for is a series of damping echoes within the time intervals of: 
\begin{eqnarray}
\boxed{\Delta t_{{\rm echo}, I }({\rm sec})
=\left\{
 \begin{matrix}
  0.2925 \pm 0.00916 & I= {\rm GW150914} \\
0.1013 \pm 0.01152 & I={\rm GW151226} \\
 0.1778 \pm 0.02789 & I={\rm LVT151012}
 \end{matrix}
 \right.}
\nonumber\\  \label{t_echo_meas}
 \end{eqnarray}
This prediction follows from combining the linear perturbation theory with the Planck-scale hypothesis, using the reported constraints on the final redshifted masses and spins of the remnant black holes \cite{TheLIGOScientific:2016pea}. As such, the reported errors are dominated by the LIGO {\it detector noise}. 

There is further {\it theoretical uncertainty} on the time-delay from merger until the first echo, given that the metric perturbations are non-linear close to the merger event:
\begin{equation}
x \equiv \frac{t_{\rm echo} - t_{\rm merger}}{\Delta t_{\rm echo}} = 1 \pm {\cal O}(1 \%).
\label{nonlinear}
\end{equation}

Our primary method was then to maximize the signal-to-noise ratio (SNR) for the echo template and see whether there is a significant peak within the predicted range given by Eq. (\ref{nonlinear}). The significance is then quantified by how often a higher peak can be found elsewhere, within a similar interval. This is best demonstrated  in Fig.  (\ref{SNR_fig}), which is similar to Fig. 4 in  \cite{Abedi:2016hgu}, but over a larger range. Indeed, without any further analysis, our main conclusions are manifest in this figure:

\begin{figure*}
  \centering
    \includegraphics[width=\textwidth]{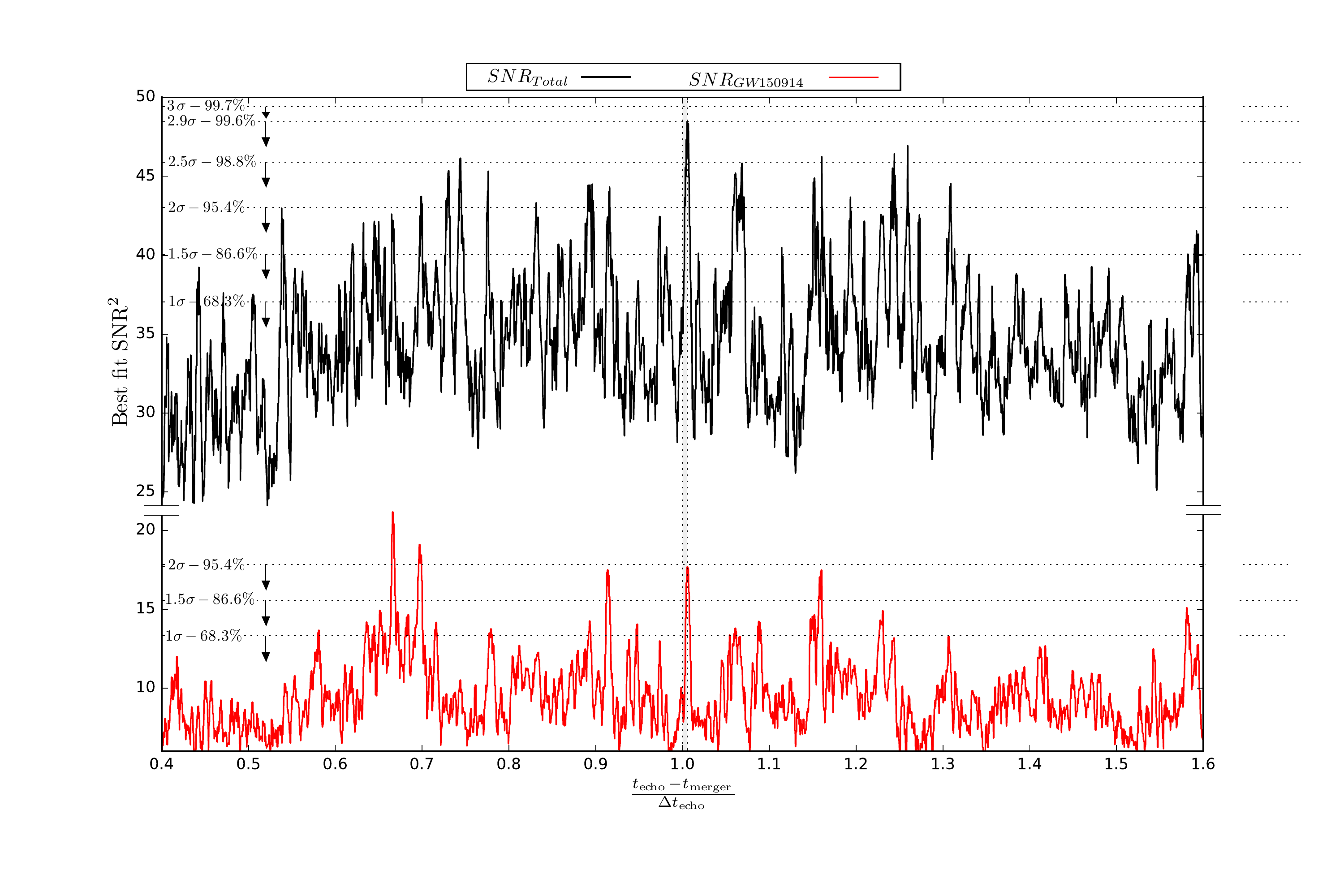}
    \caption{Same as Fig. 4 in \cite{Abedi:2016hgu}, but over an extended range of $x=\frac{t_{\rm echo} - t_{\rm merger}}{\Delta t_{\rm echo}}$. The SNR peaks at the predicted value of $x=1$ have 2.0$\sigma$ and 2.9$\sigma$ significance, for GW150914 and combined events respectively. }
 \label{SNR_fig}
\end{figure*}

For both GW150914 (the most significant reported LIGO event) and combined data from all three events, there are ubiquitous peaks within 0.54\% of $x = 1$, which is shown by the vertical grey bar (The width of the grey bar is the distance of the peaks from $x=1$). For GW150914, the significance is 2$\sigma$ (or a p-value of 5\%), meaning that comparable SNR peaks (from random noise) can be found within $\Delta x \simeq 0.0054/0.05 = 0.11$, as can be seen with other peaks at $x\simeq 0.91$ and $1.16$.  For the combined events, the significance is 2.9$\sigma$ (or a p-value of 1/270), i.e. comparable peaks can only be found within $\Delta x = 0.0054 \times 270 = 1.46$, which is also demonstrated in Fig. (\ref{SNR_fig}) as no higher peak can be seen within an interval of $\Delta x =1.2$. 

\begin{figure}
  \centering
    \includegraphics[width=0.5\textwidth]{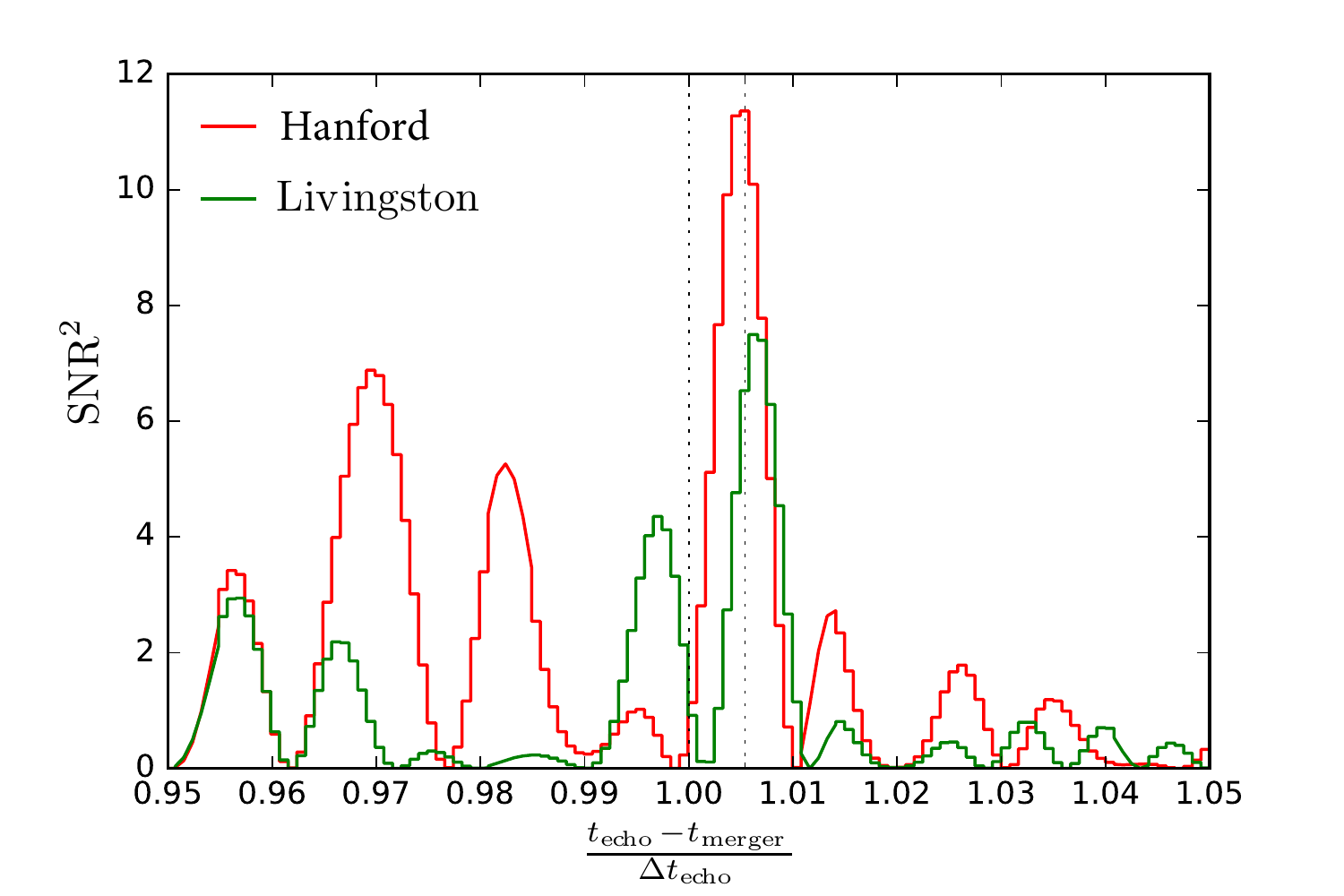}
    \caption{SNR$^{2}$ near the expected time of merger echoes (Eq. \ref{t_echo_meas}) for GW150914 in Hanford (red) and Livingston (green) detectors. Interestingly, their SNR ratio $2.74/3.37 = 0.81$ is comparable to the SNR ratio for the main event $13.3/18.6 = 0.72$. Note that, unlike Fig. (\ref{SNR_fig}), here we have fixed the echo parameters to their best fit values for combined detectors.}
 \label{Hanford_Livingston}
\end{figure}

 Indeed, the fact that the highest SNR peak in Fig. (\ref{SNR_fig}) is within $x-1=0.054$ of the theoretical prediction, while no higher peaks exists within a range of $\Delta x \gtrsim 1$ is a clear indication that this is unlikely to be mere coincidence. Even for GW150914 alone, which has a lower significance, after applying the time-delay due to finite speed of light, both Hanford and Livingston detectors see simultaneous SNR peaks near $x = 1$ (Fig. \ref{Hanford_Livingston}), which sounds very unlikely due to random chance.

With this introduction, let us now address the specific criticisms raised by Ashton et al. \cite{Ashton:2016xff}:

\begin {enumerate} 

\item Ashton et al. point out that we find a slightly higher SNR$_{\rm best}$ for echoes in LVT151012, compared to GW150914, even though the SNR for the main event is lower by a factor of 2.4.  Is this surprising?

In fact, this is expected as constraints on final mass and spin of LVT151012 are significantly worse than GW150914. As a result, the relative error on $\Delta t_{\rm echo}$ is 5 times higher for LVT151012, compared to GW150914. This leads to larger values of SNR$_{\rm best}$ across the board, as we are searching a larger region of parameter space. This, however, does not necessarily lead to increased significance, as the same would be true for all values of $x$. 

If there was no real echo signal in LVT151012 and GW151226, adding them to GW150914 would only dilute the significance of the peak near $x=1$. The fact that the opposite happens suggests that, in spite of larger variations in SNR  due to higher uncertainty in $\Delta t_{\rm echo}$, there is still significant enhancement in SNR near $x=1$ . 

We should also caution about comparing the significance of the echoes with that of the merger events, as they have very different frequency structures (see Fig. \ref{Match_frequency}) leading to different SNR ratios, especially given the non-trivial frequency dependence of the LIGO detector noise. 

Finally, we should warn about over-interpreting our quoted significances. Even though we gain comparable evidence for echoes by including LVT151012 and GW151226, i.e. $ 2^2+2^2 \simeq 2.9^2$, it doesn't mean that they have the same significance: A $2\sigma$ peak could be a 1-$\sigma$ fluctuation of a 1-$\sigma$ or a 3-$\sigma$ underlying signal. 

For completeness, the individual amplitudes of the best joint fit are listed in Table (\ref{table_1}). We note that, even though best fit SNR's are comparable for the three events, the errors on the amplitude: $\Delta A$ = $A_{\rm best}$/SNR$_{\rm best}$ is much smaller for GW150914, given that $A_{\rm best}$ is the smallest. Therefore, as expected, GW150914 which is the most significant of the 3 LIGO events, would also dominate the combined constraint on the echo amplitude. 

\begin{table}
\begin{center}
\begin{tabular}{ |c|c|c|c| }
\hline
\ \  & GW150914 & GW151226 & LVT151012 \\
 \hline
$|A_{\rm best, I}|$ & 0.091 & 0.33 & 0.34 \\ 
\hline
SNR$_{\rm best, I}$ & 4.13 & 3.83 & 4.52\\
\hline
\end{tabular}
\caption{ The best fit SNR's and amplitudes of individual events, for our joint echo template fit to the three events (see \cite{Abedi:2016hgu} for details). }\label{table_1}
\end{center}
\end{table}

\begin{figure}
  \centering
    \includegraphics[width=0.5\textwidth]{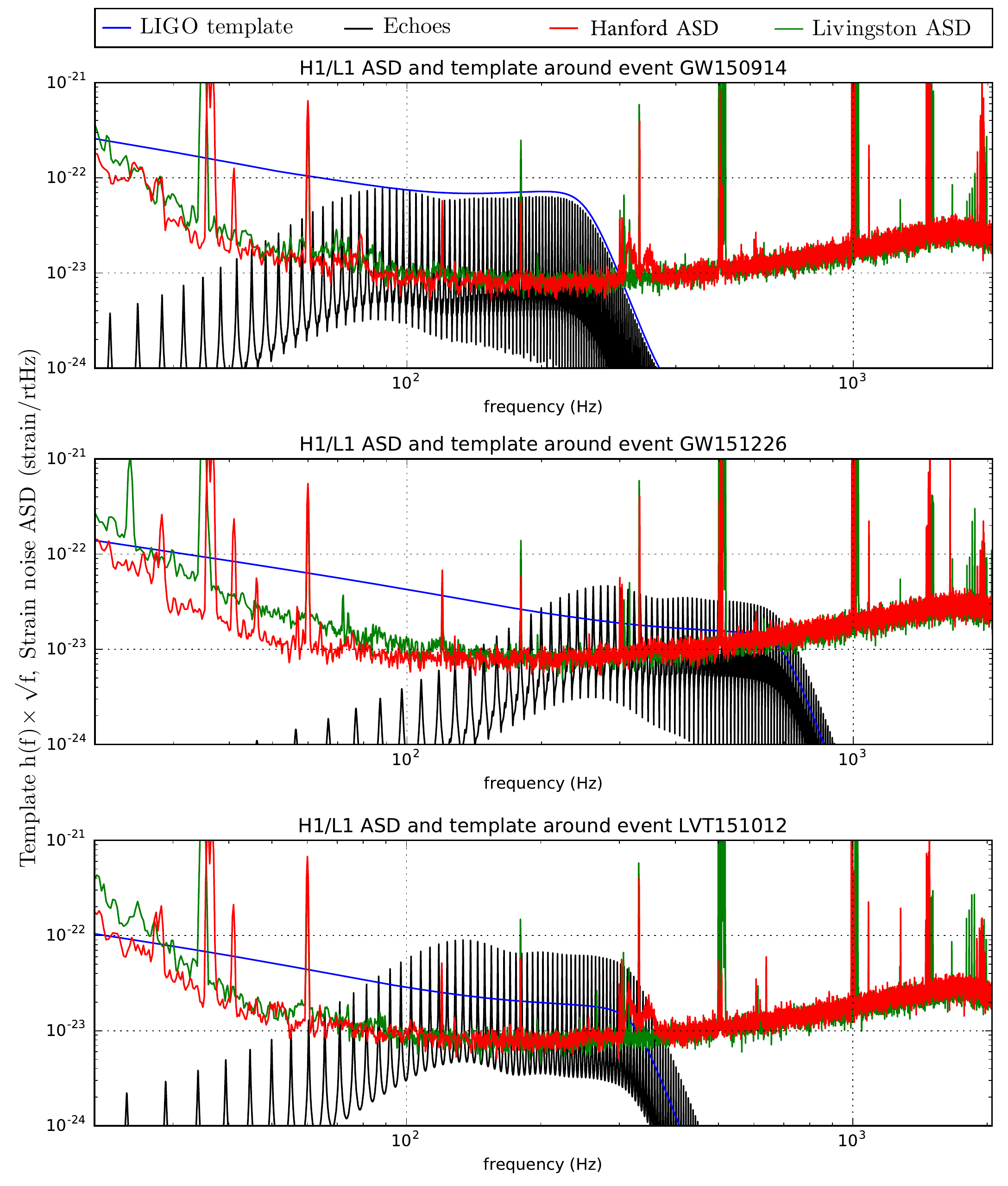}
    \caption{Best fit templates for LIGO main events and echoes (using the joint best fit described in \cite{Abedi:2016hgu}), in Fourier space (similar to Fig. 3 in \cite{Abedi:2016hgu}). The amplitude spectral distribution (ASD) for each detector is shown for comparison.  }
 \label{Match_frequency}
\end{figure}

\item Ashton, et al. worry that railing up of the best-fit SNR values near the boundary of the parameter range, particularly the damping factor $\gamma_{\rm best}=0.9$, might pose a problem for our analysis (an issue that we discussed at length in \cite{Abedi:2016hgu}) . This indeed would be the case if the goal was to measure these parameters. However, that has not been our goal, as the parameters only quantify a toy model for the echoes. The  goal was rather to find whether the best-fit toy model, within the parameter range, is consistent with random noise. As we discussed in the introduction, we find that has a probability of $<$ 1\%. 

As we argue in \cite{Abedi:2016hgu}, rather than pushing the parameters of a toy model to their extremes, in our opinion, it will be much more fruitful to find more physical echo templates. 

\item Perhaps the most serious objection of Ashton, et al. concerns our estimation of significance, or false-detection probability (p-value). 
As we outlined in the introduction, it is already clear from Fig. (\ref{SNR_fig}) that the p-value for our SNR peak near $x=1$ should be $\lesssim 0.1$  and $\lesssim 0.01$, for GW150914 and combined events, respectively.

 The main criticism of Ashton, et al. stems from us quantifying our p-values by considering how often random intervals of size $\Delta x =0.0054$ have an SNR bigger than the peaks we observe at $x=1.0054$, while we should have actually allowed for different choices of $\Delta x$. This would depend on the prior for $\Delta x$: the larger the prior, the the higher would be the p-value. 
 
 \begin{figure}[t]
  \centering
    \includegraphics[width=0.5\textwidth]{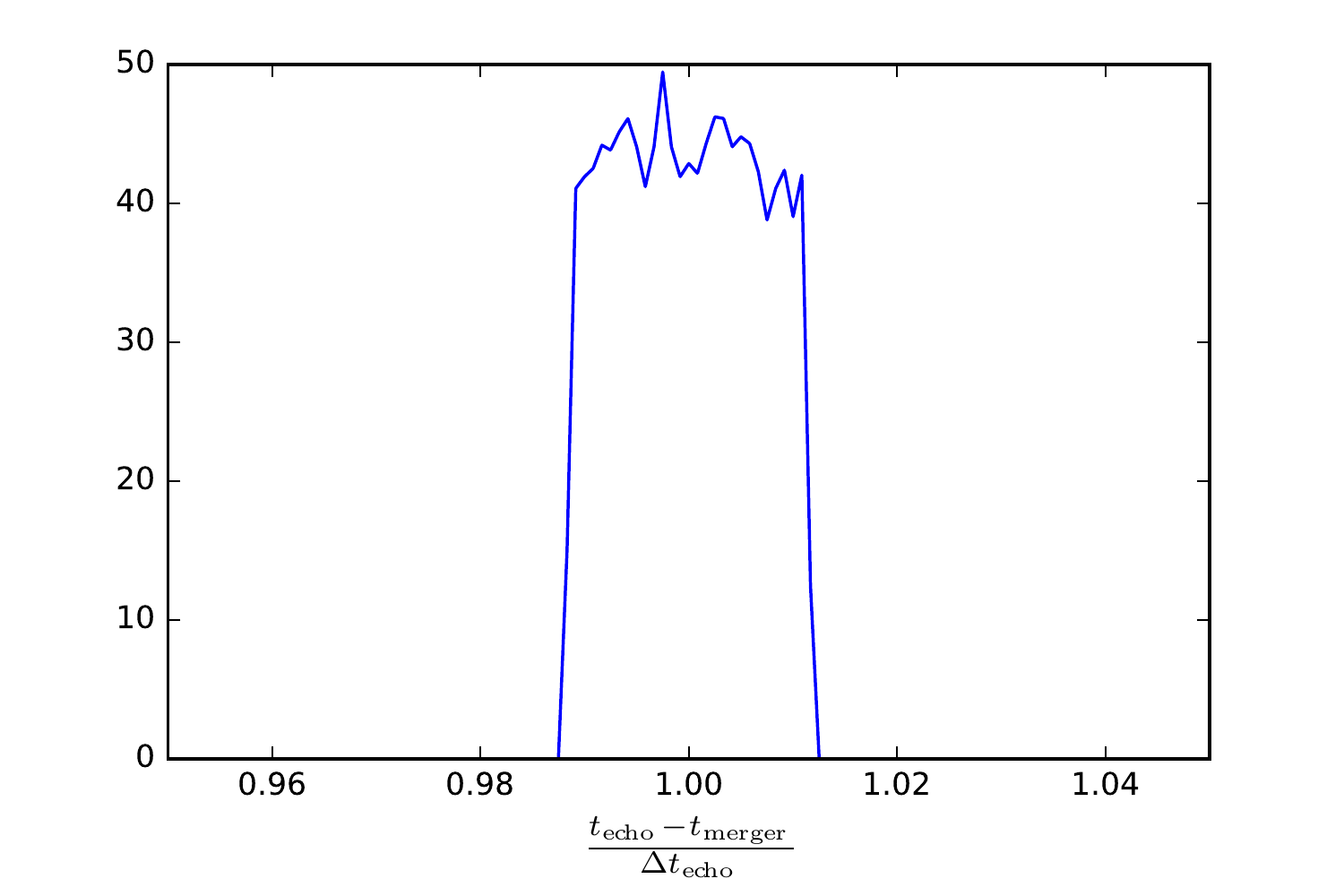}
    \caption{Resulting prior distribution on  $x=\frac{t_{\rm echo} - t_{\rm merger}}{\Delta t_{\rm echo}}$, assuming a random phase for the echo template. }
 \label{echo_prior}
\end{figure}
 
 However, we already have a decent idea about this prior from Eq. (\ref{nonlinear}) which suggests $\Delta x = {\cal O}(0.01)$, not far from what we used. We can get a more concrete handle on this prior by assuming that the echo template acquires a random phase (with respect to the main event) due to nonlinear propagation effects. Figure (\ref{echo_prior}) shows the resulting prior on $\Delta x$, which we find by replacing the data in our SNR computation (for GW150914) by the echo template with a random phase, and finding the position of the peak.  This results in a near top-hat prior with $-0.01 <\Delta x <0.01$ (an interval of $0.02$ rather than $0.0054$), which slightly increases the p-value to 0.011 (or significance of 2.54$\sigma$). 
 
Yet another way to quantify the significance would be to define a ``loudness'' function which averages the maximum likelihood for the echoes with a gaussian prior $x =1 \pm \sigma_{\rm echo}$, i.e. :
\begin{equation}
L(x,\sigma_{\rm echo}) \equiv \int \exp\left[{\rm SNR}_{\rm{total}}^{2}(x') \over 2\right] \times \frac{\exp\left[-\frac{(x-x')^{2}}{2\sigma^2_{\rm echo}}\right]}{\sqrt{2\pi \sigma^2_{\rm echo}}} dx'.\label{prior}
\end{equation}
\begin{figure}[t]
  \centering
    \includegraphics[width=0.5\textwidth]{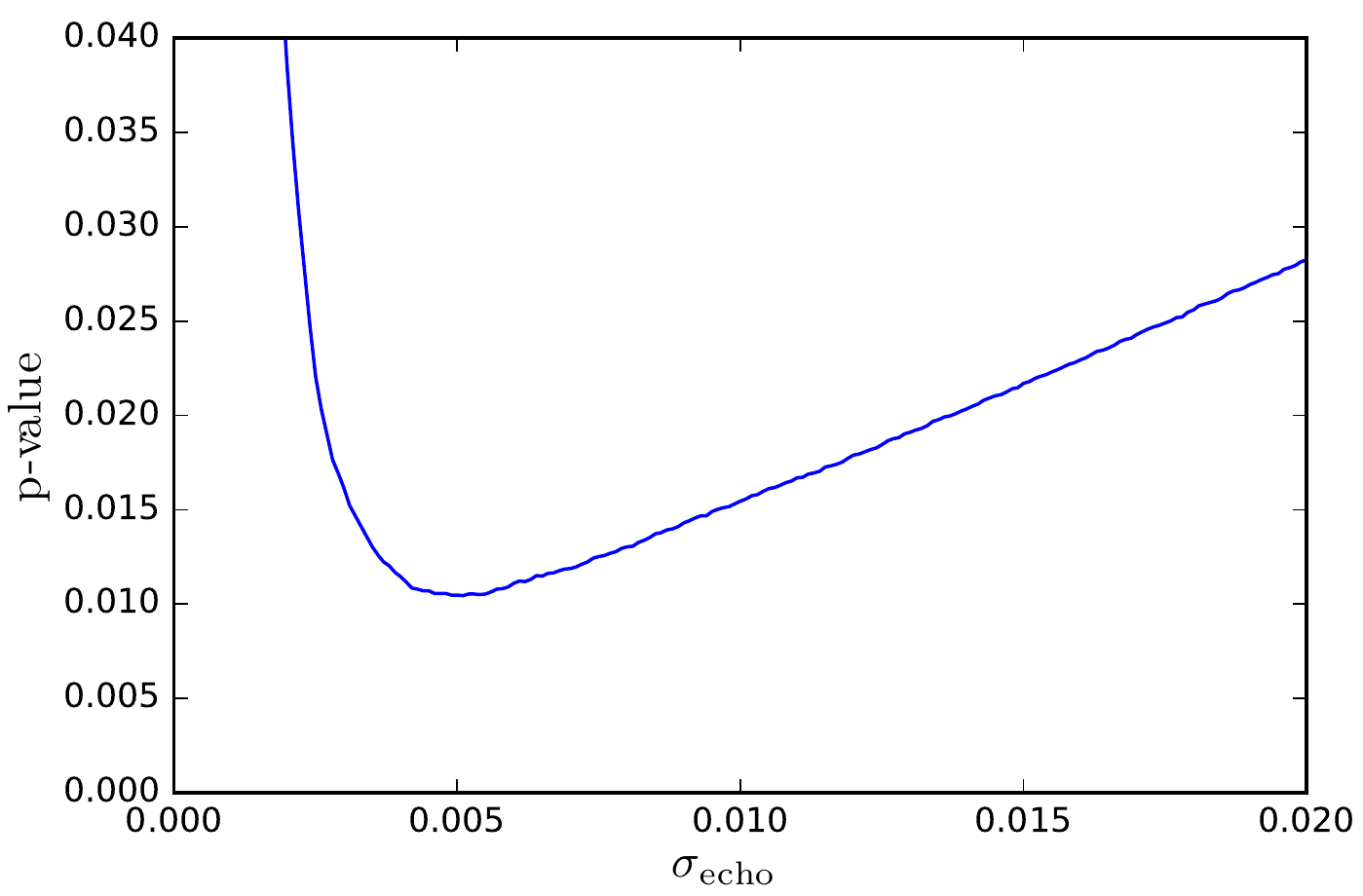}
    \caption{An alternative false detection probability (p-value) as a function of uncertainty in $t_{\rm echo}$ defined in Eq. (\ref{prior}). }
 \label{p_value}
\end{figure}
We again use the LIGO data stream within the range 9-38 $\times  \overline{\Delta t}_{\rm{echo},I}$ after the merger event, to quantify how often $L(x,\sigma_{\rm echo})$ exceeds $L(1,\sigma_{\rm echo})$, for a given $\sigma_{\rm echo}$. This plotted in Fig. (\ref{p_value}), and provides an alternative p-value (or probability of false detection). This is also minimized at $\sigma_{\rm echo} \simeq 0.5\%$, with p-value of $0.01$ (or significance of $2.6\sigma$).

\item Ashton, et al. claim that our reported significance of $2.9\sigma$ based on p-value of $3.7 \times 10^{-3}$  corresponds to $2.7\sigma$ with a one sided significance convention. We disagree: we already consider one sided significance in our p-value estimation, since we just consider the absolute value of the SNRs. 

\item Ashton et al. are concerned that the range 9-38 $\times  \overline{\Delta t}_{\rm{echo},I}$ after the merger event, which we use to quantify false detection probability, might be contaminated by the echoes and somehow affect our significance estimation. Firstly, this is unlikely, as the evidence for echoes remains marginal and nearly all LIGO data (away from the merger event) is dominated by noise. Secondly, p-value quantifies the probability of null hypothesis, i.e. how often you see the echoes, assuming that there are none. As such, to find p-value one should assume that LIGO data, away from the main event, is pure noise and use that to quantify detection probability, which was what we did.  Therefore, we find this criticism ill-founded. 

Ashton et al. further advocate using larger stretches of LIGO data (which is publicly available) to define p-value more precisely. While this is in principle correct, LIGO noise is known to significantly vary and be very non-gaussian over long time-scales (see Fig's 14-15 in \cite{Martynov:2016fzi}), which makes the interpretation of p-value ambiguous. The 9-38 $\times  \overline{\Delta t}_{\rm{echo},I}$ interval used is quite adequate to quantify the p-value for our signal, as otherwise we would see a sharp cut-off in our SNR cumulative distribution (Fig. 5 in \cite{Ashton:2016xff}).

\end{enumerate}

To conclude, while the authors of \cite{Ashton:2016xff} have raised important questions about our tentative evidence for Planck-scale structure near black hole horizons \cite{Abedi:2016hgu}, we believe they do not affect our conclusions significantly. We have provided a careful assessment of these issues, along with various quantitative and qualitative arguments for why the false detection probability remains less than $1\%$ (or $>2.6\sigma$ significance).  These will be also further discussed and expanded in an upcoming revision of our original arXiv submission. Of course, the real test will be whether this evidence can be reproduced in independent analyses by other groups, and/or using other merger events.




\begin{acknowledgments}

{\em Acknowledgments}---  We  thank Vitor Cardoso and Luis Lehner for helpful comments and discussions. We also thank the authors of \cite{Ashton:2016xff} for their very constructive feedback, as well as providing us with fruitful holiday activities! 

\end{acknowledgments}
\bibliography{Echoes_Holiday}

\end{document}